\documentclass[%
superscriptaddress,
preprint,
 amsmath,amssymb,
 aps,
]{revtex4-2}

\usepackage{graphicx}
\usepackage{dcolumn}
\usepackage{bm}
\usepackage{lineno}
\usepackage[normalem]{ulem}

\begin{document}

\title{Strain-modulated anisotropic electronic structure in superconducting RuO$_2$ films}

\author{Connor A. Occhialini}
\email{caocchia@mit.edu}
\affiliation{Department of Physics, Massachusetts Institute of Technology, Cambridge, MA 02139, USA.}

\author{Luiz G. P. Martins}
\affiliation{Department of Physics, Massachusetts Institute of Technology, Cambridge, MA 02139, USA.}

\author{Shiyu Fan}
\affiliation{National Synchrotron Light Source II, Brookhaven National Laboratory, Upton, NY 11973, USA.}

\author{Valentina Bisogni}
\affiliation{National Synchrotron Light Source II, Brookhaven National Laboratory, Upton, NY 11973, USA.}

\author{Takahiro Yasunami}
\affiliation{Department of Applied Physics, University of Tokyo, Tokyo 113-8656, Japan.}

\author{Maki Musashi}
\affiliation{Department of Applied Physics, University of Tokyo, Tokyo 113-8656, Japan.}

\author{Masashi Kawasaki}
\affiliation{Department of Applied Physics, University of Tokyo, Tokyo 113-8656, Japan.}

\author{Masaki Uchida}
\affiliation{Department of Physics, Tokyo Institute of Technology, Tokyo 152-8550, Japan.}

\author{Riccardo Comin}
\email{rcomin@mit.edu}
\affiliation{Department of Physics, Massachusetts Institute of Technology, Cambridge, MA 02139, USA.}

\author{Jonathan Pelliciari}
\email{pelliciari@bnl.gov}
\affiliation{National Synchrotron Light Source II, Brookhaven National Laboratory, Upton, NY 11973, USA.}

\begin{abstract}
{The binary ruthenate, RuO$_2$, has been the subject of intense interest due to its itinerant antiferromagnetism and strain-induced superconductivity. The strain mechanism and its effect on the microscopic electronic states leading to the normal and superconducting state, however, remain undisclosed. Here, we investigate highly-strained epitaxial (110) RuO$_2$ films using polarization-dependent oxygen K-edge X-ray absorption spectroscopy (XAS). Through the detection of pre-edge peaks, arising from O:$2p$ - Ru:$4d$ hybridization, we uncover the effects of epitaxial strain on the orbital/electronic structure near the Fermi level. Our data show robust strain-induced shifts of orbital levels and a reduction of hybridization strength. Furthermore, we reveal a pronounced in-plane anisotropy of the electronic structure along the $[110]/[1\bar{1}0]$ directions naturally stemming from the symmetry-breaking epitaxial strain of the substrate. The $B_{2g}$ symmetry component of the epitaxially-enforced strain breaks a sublattice degeneracy, resulting in an increase of the density of states at the Fermi level ($E_F$), possibly paving the way to superconductivity. These results underscore the importance of the effective reduction from tetragonal to orthorhombic lattice symmetry in (110) RuO$_2$ films and its relevance towards the superconducting and magnetic properties.}
\end{abstract}

\date{\today}
\maketitle

The recent discovery of superconductivity in RuO$_2$, the simplest ruthenate, has spurred wide interest in the community \cite{Uchida2020,Ruf2021}. Although early calculations based on Migdal-Eliashberg theory predicted a superconducting phase of conventional origin \cite{Glassford1994,Lin2004}, superconductivity has evaded detection in bulk single crystals despite significant efforts to measure resistivity down to sub-Kelvin temperatures \cite{Lin2004}. Only recently, through the controlled growth and stabilization of highly-strained, epitaxial RuO$_2$ films of high quality, superconductivity has been observed with $T_c \sim 1.8$ K \cite{Uchida2020,Ruf2021}.

The precise mechanism for this strain-induced superconductivity is a subject of debate, although a conventional (phonon-mediated) pairing mechanism is supported by first-principles calculations and basic arguments of BCS theory \cite{Uchida2020,Ruf2021,Glassford1994}. From this perspective, there are two main potential sources for the emergence of a superconducting ground state: (i) a modification of the phonon spectrum producing an enhancement of electron-phonon coupling (EPC); and (ii) an increase of the density of states at the Fermi level, providing sufficient energy gain for the opening of the superconducting gap. 
Here, we will address point (ii) and investigate the evolution of the unoccupied density of states, their symmetry, and Ru-O hybridization in strained RuO$_2$ films through oxygen $K$-edge X-Ray absorption spectroscopy (XAS) in superconducting and non-superconducting samples. 

Oxygen $K$-edge XAS (O:$1s$ $\to$ O:$2p$) at the pre-edge allows a measurement of the Ru:$4d$ density of states through Ru:$4d$/O:$2p$ hybridization \cite{Frati2020}. Our experiments reveal a significant evolution of the Ru:$t_{2g}$ states as a function of strain in RuO$_2$ epitaxial films. XAS also permits orbital-selectivity through polarization selection rules, allowing us to clearly isolate the effects of the anisotropic strain induced by the epitaxial growth. Additional low-energy features in the XAS profile are observed, corresponding to a strain-modulated electronic structure of unoccupied states near the Fermi level ($E_F$) tightly connected to the large $c$-axis compression and the anisotropic strain in the $ab$-plane. These observations reflect the key role of the broken tetragonal symmetry of the rutile lattice in epitaxial films. The confluence of these engineered anisotropic strain states results in an increase of the density of states near $E_F$, an essential ingredient for unraveling the origins of superconductivity.

\begin{figure}[htb]
\includegraphics[width=\columnwidth]{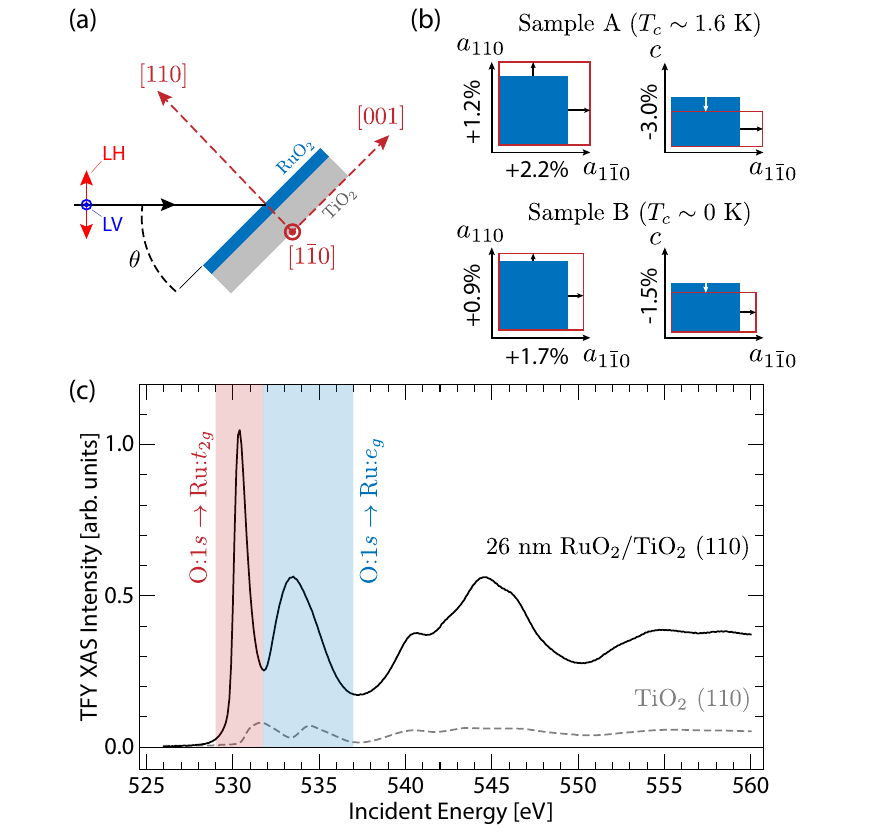}
\caption{(a) Crystallographic orientation, scattering geometry and polarization states (LH/LV) used for the experiments. (b) Schematic of the strain-states for the samples: superconducting film (Sample A) and non-superconducting film (Sample B). (c) Oxygen K-edge absorption spectrum in TFY for Sample A at $\theta = 15$ degrees and LH polarization compared to the absorption spectrum of the bare TiO$_2$ substrate. }
\label{fig:fig1}
\end{figure}

To study the effect of strain on the electronic structure, we performed measurements on two similarly-prepared thin films of RuO$_2$. A highly-strained, superconducting 26 nm RuO$_2$ (110)/TiO$_2$ (110) film ($T_{c} \sim 1.6$ K) (labeled Sample A) is compared to a non-superconducting, partially strain-relaxed 30 nm RuO$_2$ (110)/TiO$_2$ (110) film (Sample B) \cite{Uchida2020}. The superconducting film is prepared on a step-terraced TiO$_2$ substrate, reducing strain relaxation and resulting in closer lattice matching \cite{Uchida2020}. The lattice strains for the two samples are summarized in Fig. \ref{fig:fig1}(b) (see Ref. \citenum{SupplementalCitation} for sample characterization). Oxygen $K$-edge XAS spectra were recorded as a function of incidence angle and incident linear polarization (linear vertical [LV] and horizontal [LH]), in Total Fluorescence Yield (TFY) at the 2-ID (SIX) beamline of NSLS-II at Brookhaven National Laboratory \cite{Dvorak2016}, using the scattering geometry depicted in Fig. \ref{fig:fig1}(a). All measurements were performed at $T = 40$ K.

\begin{figure}[htb]
\includegraphics[width=\columnwidth]{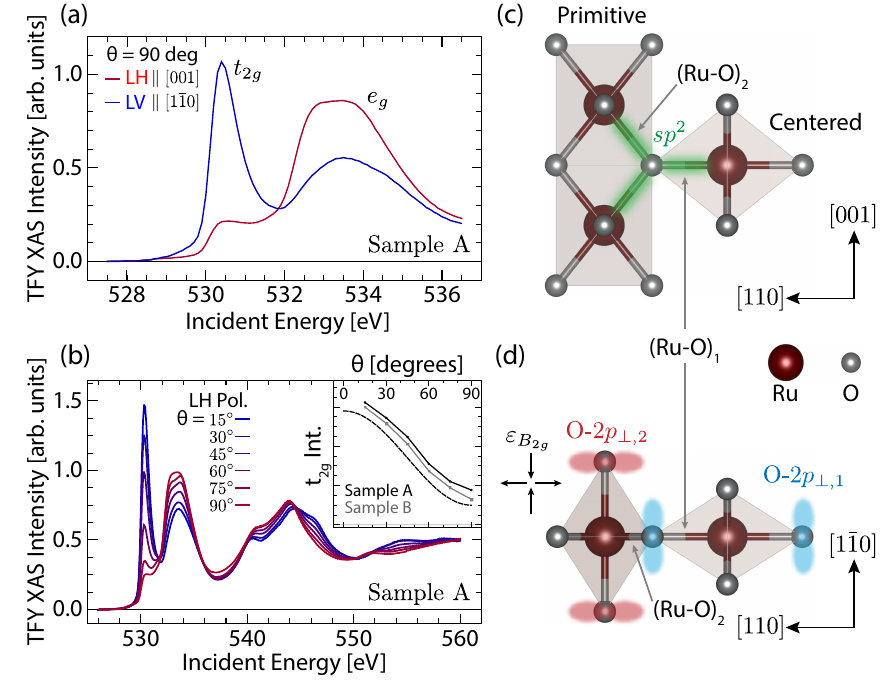}
\caption{(a) Normal incidence linear dichroism O K-edge X-ray Absorption Spectroscopy (XAS) spectra of Sample A. (b) Angular dependent LH-polarized XAS spectra on Sample A. The inset in (b) shows the angular dependence of the integrated $t_{2g}$ peak intensity for Samples A (black dots) and B (grey dots), showing a cosinusoidal dependence with a 6-fold suppression of the $t_{2g}$ pre-edge peak intensity for the out-of-plane polarization ($\epsilon \parallel [001]$). The dashed line in the inset is a guide to the eye for $\cos^2(\theta)$ angular dependence. (c) Local trigonal coordination of oxygen with proximal Ru ions forming a dominant $sp^2$ type bond interacting primarily with the Ru $e_g$ states (green). The remaining O-$2p_{\perp}$ orbitals are denoted in (d). These bonding states result from the distinct edge/corner sharing connectivity of the RuO$_6$ octahedra along the $c$ and $a$/$b$ axes, respectively, as shown in (c)/(d).  Also indicated in (c) are the two distinct types of bonds, the purely in-plane oriented (Ru-O)$_1$ bond and the (Ru-O)$_2$ bond with finite projection along the $c$-axis.}
\label{fig:fig2}
\end{figure}

In Fig. \ref{fig:fig1}(c) we report a selection of XAS spectra taken on an RuO$_2$ (110)/TiO$_2$ (110) film as well as on the bare TiO$_2$ (110) substrate. From our XAS profiles we can disentangle the signals originating in the RuO$_2$ epitaxial films and the underlying TiO$_2$ substrate. The pre-edge regions correspond to the hybridization peaks associated to the Ru:$t_{2g}$ ($E_i \simeq 530.1$ eV) and the Ru:$e_g$ ($E_i \simeq 533.3$ eV) states, which are active in $\pi$- and $\sigma$-bonding with the oxygen $2p$ states, respectively \cite{Hu2000,Occhialini2021}. In Fig. \ref{fig:fig2}(a) we report the linear dichroism at normal incidence on Sample A (superconducting). The two XAS spectra correspond to incident polarization LV $\parallel [1\bar{1}0]$ and LH $\parallel [001]$. The $t_{2g}$ pre-edge intensity is strongly suppressed for polarization along the [001] axis, which suggests a predominance of $\pi$-bonding O-$2p$ orbitals oriented within the $ab$-plane. This is further elucidated by the angular dependence in LH polarization reported in Fig. \ref{fig:fig2}(b). At normal incidence, LH $\parallel [001]$ and as the sample is rotated, the polarization projects to LH $\parallel [110]$ (the surface normal of the film). Plotting the integrated $t_{2g}$ intensity versus incident angle, we find a cosinusoidal dependence with a six-fold increase of intensity when the polarization is oriented within the (001) plane as reported in the inset of Fig. \ref{fig:fig2}(b).

This result may be understood by the reduced $\pi$-bonding strength of the $t_{2g}$ states, as discussed in the context of bulk RuO$_2$ and isostructural rutile compounds \cite{Sorantin1992,Glassford1994,Goodenough1971a,Stagarescu2000,Occhialini2021,Kahk2014,Ping2015}. For each oxygen atom, there is a nearly-trigonal coordination of Ru atoms connecting the two Ru sublattices (at the centered and primitive lattice positions of the body-centered unit cell, respectively) as denoted in Fig. \ref{fig:fig2}(c). This geometry favors an $sp^2$-type bonding scheme, active through $\sigma$-type bonds with the $e_g$ states. For each such configuration, this leaves one orbital perpendicular to the trigonal coordination plane on each of the differently oriented oxygen sites as the most active in $\pi$ bonding, and denoted O-$2p_{\perp,{1}}$ and  O-$2p_{\perp,{2}}$ in Fig. \ref{fig:fig2}(d). The amplitude for $1s \to 2p$ dipole transitions is $\propto \cos^2(\theta)$ where $\theta$ is the angle between the incident polarization and the principle axis of the $p$-orbital. Thus, our results confirm the dominant role of O:$2p$ orbitals oriented within the $ab$-plane for $\pi$-bonding with the $t_{2g}$ states \cite{Frati2020,Das2018}. Overall, similar results were obtained for the non-superconducting Sample B \cite{SupplementalCitation}.

\begin{figure}[htb]
\includegraphics[width=\columnwidth]{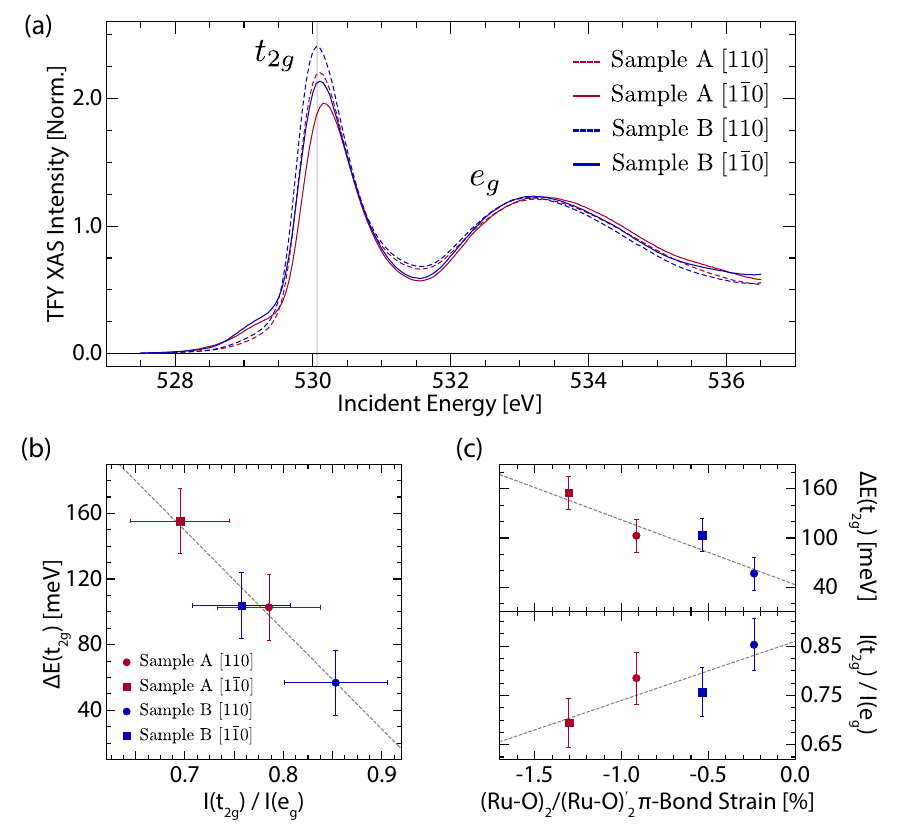}
\caption{(a) Comparison of the XAS spectra measured along the nominally equivalent $[110]$ and $[1\bar{1}0]$ directions in Samples A and B. Spectra have been normalized by the integrated O:$2p$ $\to$ Ru:$e_g$ spectral weight. (b) Shift of the $t_{2g}$ energy $\Delta E(t_{2g})$ plotted against the ratio $I(t_{2g})/I(e_g)$ of the integrated $t_{2g}$ and $e_g$ transition spectral weights reveals a linear correlation. (c) Plots of $\Delta E(t_{2g})$ (top) and $I(t_{2g})/I(e_g)$ (bottom) against the relative (Ru-O)$_2^{(')}$ bond strain. The (Ru-O)$_2$ bond length is used for $\epsilon \parallel [1\bar{1}0]$ measurements (squares) and the (Ru-O)$_2'$ bond is used for $\epsilon \parallel [110]$ (circles) (see text and Ref. \onlinecite{SupplementalCitation} for details). In (b,c), red/blue color denotes Sample A/B, respectively, and squares/circles indicate measurements with polarization along the $[1\bar{1}0]/[110]$ directions, respectively.}
\label{fig:fig3}
\end{figure}

We now consider the effects of epitaxial strain on the Ru $t_{2g}$ orbital states near the Fermi level. In Fig. \ref{fig:fig3}(a), we report absorption measurements with incident polarization along $[110]$ and $[1\bar{1}0]$ in both Sample A and B, probed with LH and LV polarizations at $\theta = 15$ degrees, respectively. The spectra are normalized to the integrated $e_g$ spectral weight. Two features of the data are remarkable: (i) a clear reduction in the ratio between the $t_{2g}$ and $e_g$ spectral weights, reported as $I(t_{2g})/I(e_g)$; and (ii) a shift of the $t_{2g}$ peak energy, indicated as $\Delta E(t_{2g}) = E(t_{2g})- 530$ eV. Plotting these quantities against each other in Fig. \ref{fig:fig3}(b) reveals a linear correlation naturally implying their coupling.  To further understand this observation, we recall that the $[110]$ and $[1\bar{1}0]$ directions experience anisotropic strain due to lattice relaxation in the out-of-plane direction of the film, as opposed to the more strict lattice matching enforced at the film/substrate interface. The antisymmetric component of this strain corresponds to a $B_{2g}$ symmetry lattice distortion [Fig. \ref{fig:fig2}(d)] that breaks the tetragonal structural symmetry in the $ab$-plane, as discussed more below. This broken symmetry, along with the difference in the $c$-axis strain between the two films, yields a selective modification of different Ru-O bond lengths \cite{Uchida2020}. In particular, the bonds identified in Fig. \ref{fig:fig2}(c,d) are split into two subsets: one set determined using the $a_{110}$ lattice parameter [(Ru-O)$_1$, (Ru-O)$_2$] and the other using $a_{1\bar{1}0}$ [(Ru-O)$_1'$, (Ru-O)$_2'$] (see Ref. \citenum{SupplementalCitation} for additional details). 

Since the $t_{2g}$ states hybridize with oxygen through $\pi$-bonds (Fig. \ref{fig:fig2}), the bonding strength will depend on bond lengths perpendicular to the principle axis of the corresponding O:$2p$ orbital probed by XAS \cite{Khomskii2014}.  Therefore, where we measure the O-$2p_{\perp,1}$ orbital with incident polarization along $[1\bar{1}0]$ [Fig. \ref{fig:fig2}(d)], we consider potential correlations between the electronic structure and the first set of bond lengths, (Ru-O)$_1$/(Ru-O)$_2$. These have in-plane projection along the $[110]$ direction and vice versa for polarization parallel to $[110]$ (see Ref. \citenum{SupplementalCitation} for a detailed discussion).  Following this scheme, we find that both $I(t_{2g})/I(e_g)$ and the $t_{2g}$ peak energy display a linear dependence on the appropriately chosen (Ru-O)$_2$/(Ru-O)$_2'$ bond-length relevant for $\pi$-bonding with the respectively probed orbital states [Fig. \ref{fig:fig3}(c)].  

In a molecular orbital picture the $t_{2g}$ states form the anti-bonding states and their energy should decrease along with a reduction of hybridization with the O:$2p$ states \cite{Khomskii2014}. This interpretation matches the observed trend of the $t_{2g}$ peak energy with respect to the (Ru-O)$_2^{(')}$ bond length [Fig. \ref{fig:fig3}(c)]. The intensity variation of the $t_{2g}$ pre-edge peaks can be due to two factors: a filling of the band affecting the number of oxygen holes; or a decrease in the hybridization strength, thereby reducing the overall O:$2p$ character of the $t_{2g}$ band \cite{Frati2020,Khomskii2014,Mirjolet2021,Fabbris2016}. The observed increase of the ratio $I(t_{2g})/I(e_g)$ with an increase of the (Ru-O)$_2$ bond length is attributable to a decrease of the hybridization (the decrease is more pronounced for the more strongly bonded $\sigma$-bonding states at the $e_g$ peak) \footnote{The intensity ratio $I(t_{2g})/I(e_g)$ plotted in Fig. \ref{fig:fig3} was corrected for an angular-dependent term for the LH polarization projection due to the finite $\theta = 15$ deg incidence angle.  The correction factor to the ratio was determined empirically from the angular-dependent data in Fig. \ref{fig:fig2}, which is close to the expected value of $\cos^2(15^\circ) \simeq 0.933$. This correction is minor and its inclusion does not effect the main results.}. These considerations provide a consistent interpretation of the experimental data in connection to the strain-effect on the Ru:$t_{2g}$ states and identifies the central role of the (Ru-O)$_2$ bond length for strain modification of the electronic structure.

As previously suggested, the emergence of $T_c$ in thin RuO$_2$ films correlates with the large $c$-axis strain which is unique to both the TiO$_2$ substrate and the (110) orientation \cite{Uchida2020}. Our results support the essential nature of the $c$-axis strain for modifying the electronic structure as well. The changes in the $t_{2g}$ levels correlate well with the (Ru-O)$_2$ bond strain, as opposed to the distinctly oriented (Ru-O)$_1$ bond [see Fig. \ref{fig:fig2}(c) and Ref. \citenum{SupplementalCitation}]. While this is the dominant effect as evidenced from the data, we note that the slight deviations from linear behavior in Fig. \ref{fig:fig3}(c) may be due to smaller contributions from the (Ru-O)$_1^{(')}$ bonds. The key distinction between the two bond types is that the apical (Ru-O)$_1$ bond is oriented purely within the $ab$ plane, and therefore its length is not susceptible to the large $c$-axis compression. Thus, these findings strongly suggest that the modifications to the orbitals near $E_F$ are tightly connected to the development of superconductivity. 

\begin{figure}[htb]
\includegraphics[width=\columnwidth]{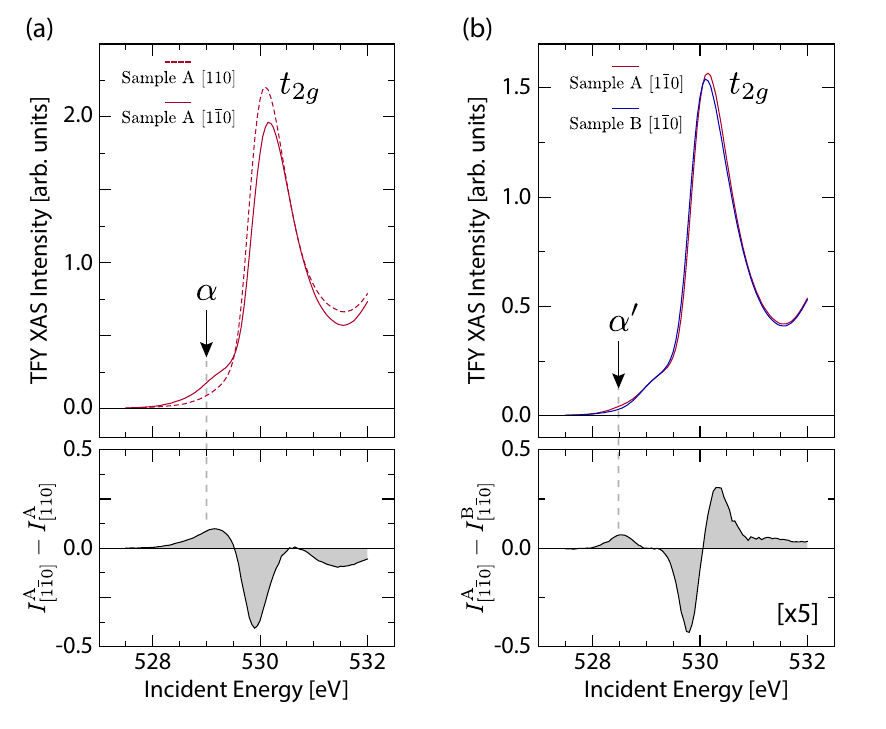}
\caption{(a) Comparison (top) and difference spectrum (bottom) between the $[110]$ and $[1\bar{1}0]$ polarized XAS spectra in superconducting Sample A. The difference spectrum highlights the emerging low-energy peak around 529.0 eV, denoted as $\alpha$. (b) Comparison (top) and difference spectrum (bottom) between the $[1\bar{1}0]$ polarized XAS spectra of Sample A and Sample B. The difference spectrum highlights the enhancement of the low-energy peak in the most strained superconducting sample. The peak extends to lower energy, down to 528.5 eV (labelled as $\alpha'$) for Sample A. The difference spectrum is multiplied by a factor of 5 in (b) for clarity.}
\label{fig:fig4}
\end{figure}

Crucially, we not only isolate these effects in distinctly strained films, but we also reveal a pronounced anisotropic strain along the $[110]/[1\bar{1}0]$ directions within each film. The electronic anisotropy in the $ab$ plane highlights the role of the broken tetragonal symmetry in epitaxial films, which is preserved in the bulk \cite{Berlijn2017,Boman1970,Rao1969}. To further elaborate on this, we investigate the lineshape of the $t_{2g}$-derived peak in more detail in Fig. \ref{fig:fig4}. Figure \ref{fig:fig4}(a) shows a close-up of the $t_{2g}$ pre-edge region for Sample A with polarization along $[110]/[1\bar{1}0]$ directions, normalized to the $e_g$ spectral weight.  Besides the aforementioned features, we detect an additional peak on the low-energy side of the $t_{2g}$ peak ($E \sim 529$ eV) only for the polarization $\epsilon \parallel [1\bar{1}0]$, denoted as peak $\alpha$ in the linear dichroism spectrum on the bottom panel of Fig. \ref{fig:fig4}(a). While XAS spectra of comparable quality on bulk RuO$_2$ are so-far undisclosed, the available data indicate an absence of this peak \cite{Hu2000,Occhialini2021}, consistently with the band structure calculations of unstrained bulk RuO$_2$ \cite{Glassford1994,Berlijn2017,Uchida2020,Ruf2021}. The new feature also appears in Sample B, only for polarization along $[1\bar{1}0]$ similarly to Sample A [Fig. \ref{fig:fig4}(b)]. Normalizing the $[1\bar{1}0]$ spectra in each sample to the $t_{2g}$ intensity (to factor out the effects of the modified hybridization) indicates that the new spectral weight is present in both samples, but it extends to lower incident energy in Sample A [denoted by $\alpha'$ in Fig. \ref{fig:fig4}(b)]. This last observation indicates an enhanced unoccupied density of states close to to the Fermi level of Sample A compared to Sample B.

The reliability of our results (Fig. \ref{fig:fig3} and \ref{fig:fig4}) was extensively confirmed through cyclical measurements in different positions on both samples demonstrating the robust energy calibration and the reproducible low-energy peaks (see Fig. S4 in Ref. \citenum{SupplementalCitation}). In the geometry used for data of Fig. \ref{fig:fig4}, the penetration depth of X-Rays is $\simeq 25$ nm \cite{Henke1993}. Thus, while the measured spectra are sensitive to the full thickness of the film, any contributions from the TiO$_2$ substrate will be heavily suppressed. Furthermore, both pristine [Fig. \ref{fig:fig1}(c)] and doped \cite{Thomas2007,Parras2013} TiO$_2$ show no absorption signal in this energetic region. The new spectral weight is therefore attributed to an intrinsic, bulk property of the strained RuO$_2$ films, and may be associated to an increase in the unoccupied O:$2p$ density of states at (and above) the Fermi threshold.

In RuO$_2$, the Fermi level occurs near a local minimum in the DOS \cite{Glassford1994,Berlijn2017,Uchida2020,Ruf2021}, which has been used to explain its structural and electronic phase stability in contrast to other rutile compounds \cite{Goodenough1971a}. The DOS, however, increases sharply on either side of $E_F$ and therefore, small shifts of the orbital levels will couple to a large increase in the DOS at $E_F$ ($\mathcal{D}(E_F)$). The low-energy signal is attributed to additional unoccupied states above $E_F$ which are induced by strain and not present in the unstrained, bulk sample. The unoccupied component of the Ru:$t_{2g}$ states form a peak centered approximately 0.8 eV above $E_F$ in the unstrained case \cite{Ruf2021, Berlijn2017, Smejkal2020, Glassford1994}. Thus, the new signal observed near 529.0 eV is near the expected Fermi level threshold, considering its relative position compared to the main $t_{2g}$ resonance close to 530.1 eV. In the simplest approximation, O $K$-edge absorption measures the unoccupied, O-$2p$ projected DOS \cite{Frati2020}; therefore, our results directly corroborate that strain effectively induces an enhancement of $\mathcal{D}(E_F)$, as revealed by the appearance of additional unoccupied states just above $E_F$. These considerations are evidenced by the concurrence of the additional peak appearing only along the most strained (in-plane $[1\bar{1}0]$) axis in each film, as well as its enhancement when going from Sample B to Sample A.

The symmetry-breaking strain enforced by the epitaxial growth conforms to a dominant $B_{2g}$ symmetry in the \textit{ab}-plane of RuO$_2$ \cite{Moriya1959,Disa2020}, corresponding to an antisymmetric strain along the $[110]$ and $[1\bar{1}0]$ directions [see Fig. \ref{fig:fig2}(d)].  This particular symmetry of strain couples distinctly to each sublattice [see Fig. \ref{fig:fig2}(c,d)] of Ru due to the relative four-fold ($C_{4z}$) rotation of the local octahedral environments whose apical directions [corresponding apical (Ru-O)$_1$ bonds] are oriented along the $[110]/[1\bar{1}0]$ directions, respectively.  Under $B_{2g}$ strain, one sublattice becomes apically compressed while the other is elongated [Fig. \ref{fig:fig2}(c,d) and \onlinecite{SupplementalCitation}]. We further remark that the O:$2p_{\perp,1/2}$ orbitals, probed with $[1\bar{1}0]/[110]$ oriented polarization, respectively, play reciprocal roles for the two sublattices. The O:$2p_{\perp,1}$ is at the planar/apical oxygen site of the primitive/centered sublattice sites and interacts with the Ru $4d$ states through the distinct (Ru-O)$_2$/(Ru-O)$_1$ bonds, respectively [see \onlinecite{SupplementalCitation}, Fig. S2 and Fig. \ref{fig:fig2} (c,d)]. Therefore, the difference in absorption spectra when selectively probing O:$2p_{\perp,1/2}$ reflects the breaking of the sublattice degeneracy, induced by the $B_{2g}$ component of the strain. The nature of this sublattice degeneracy breaking may be responsible for a substantial enhancement of $\mathcal{D}(E_F)$ compared to films with $(100)$ orientation, through an effective transfer of charge between bands derived from the two sublattices, yielding a corresponding shift of the Fermi level to a point with higher DOS while conserving the overall electron filling.

Our experiments directly elucidate the essential nature of both the large $c$-axis compression \cite{Uchida2020, Ruf2021} as well as the anisotropic strain in the $ab$ plane which couples strongly to an anisotropic electronic structure near $E_F$. The coexistence of these anisotropic strain states yields a significant increase in $\mathcal{D}(E_F)$ for orbitals oriented within the plane of the film, which is of direct relevance for explaining the strain-induced superconductivity from the perspective of the BCS theory. The origin of this $\mathcal{D}(E_F)$ increase may be effectively linked to an `anisotropic' doping of holes into the $\pi$-bonding band derived from the Ru:$t_{2g}$ states which is nearly fully filled in the bulk \cite{Berlijn2017,Goodenough1971a,Uchida2020,Ruf2021}. Indeed, such a scenario has been hypothesized to yield superconductivity by Goodenough several decades ago \cite{Goodenough1971a}. The observed enhancement of $\mathcal{D}(E_F)$ also trends well with the appearance of superconductivity [Fig.~\ref{fig:fig4}(b)], as well as with the increased epitaxial $B_{2g}$ strain in Sample A in comparison to sample B (Table S2\cite{SupplementalCitation}). Our data corroborate recent ARPES results that report an increased $\mathcal{D}(E_F)$ as seen from the perspective of the occupied states \cite{Ruf2021}, whereas our XAS measurements reveal a complementary effect in the unoccupied states (Fig. \ref{fig:fig4}). Beyond this, we demonstrate the definitive O:$2p$ bonding character of the states near $E_F$, supporting the notion that the $\pi$-bonding $t_{2g}$ band shifting above $E_F$ is the source of this effect, rather than the non-bonding $d_{x^2-y^2}$ states \cite{Ruf2021} which form a sharp peak in the DOS nearly 1 eV below $E_F$ in the unstrained case \cite{Goodenough1971a,Berlijn2017,Ping2015}. In addition, our results reveal that the increase of $\mathcal{D}(E_F)$ is highly anisotropic and tightly connected to the tetragonal symmetry-breaking induced by the epitaxial strain state unique to (110) films. These observations will serve as an essential constraint for future theoretical work to determine the complete electronic structure in strained RuO$_2$, as well as for assessing the role of alternative interactions relevant for superconductivity, such as renormalized electron-phonon coupling \cite{Uchida2020, Ruf2021}.

As we have demonstrated, the in-plane ($[1\bar{1}0]$) and out-of-plane ($[110]$) absorption spectra [Fig. \ref{fig:fig4}(a)] project out the distinct electronic structure associated to each Ru sublattice and the differences between the two also suggest a pronounced breaking of the nominal Ru sublattice degeneracy in the regime near $E_F$ relevant for transport and superconductivity. In this vein, RuO$_2$ films are not only unique for their superconductivity, but also for itinerant antiferromagnetism \cite{Zhu2019, Berlijn2017} and their promise for spintronics applications \cite{Bai2021,Feng2020,Smejkal2020,Smejkal2022, Gonzalez-Hernandez2021}. This includes the emergence of an anomalous Hall conductivity uniquely in (110) films that is absent in (100) orientation \cite{Smejkal2020,Feng2020}. Itinerant antiferromagnetism in RuO$_2$ is connected to a Fermi surface instability \cite{Zhu2019,Berlijn2017,Smejkal2020}, which may be strongly affected by strain-induced changes modifying the electronic structure. For the magnetic order proposed in bulk RuO$_2$, the centered and primitive Ru sites form the sublattices of the antiferromagnetic structure with local magnetic moments nearly parallel and antiparallel to the $[001]$ direction, respectively \cite{Zhu2019,Berlijn2017}. Therefore, the breaking of the structural sublattice degeneracy under a $B_{2g}$ strain in (110) films naturally implies a breaking of the compensation between the collinear antiferromagnetic sublattices, potentially yielding weak ferromagnetism \cite{Dzyaloshinsky1958}.
Such a piezomagnetic effect in rutile antiferromagnets is of historic and recent interest \cite{Moriya1959,Disa2020}. While more work is needed to confirm this scenario, we want to underscore that the symmetry-breaking $B_{2g}$ strain is also relevant for the magnetic properties, as structural and magnetic symmetries are naturally intertwined in RuO$_2$ \cite{Smejkal2020, Feng2020, Bai2021, Smejkal2022}.

In conclusion, we have presented a detailed polarization-dependent oxygen K-edge XAS study in epitaxially-strained films of RuO$_2$. Polarization selection rules allow a determination of the symmetry of the oxygen orbitals responsible for bonding with the $t_{2g}$ states. Further measurements on differently strained samples reveal a robust evolution of the spectra as a function of the strain state, including orbital energy shifts and modified hybridization strength which are associated to the large $c$-axis compression in RuO$_2$ films. In addition, our measurements uncover evidence for an increased density of states at the Fermi level, connected to the breaking of sublattice symmetry concomitant with the presence of superconductivity. These results underscore the importance of anisotropic strain in (110) RuO$_2$ films and its direct implications for their unique superconducting and magnetic properties.

\textit{Acknowledgements.} This work was supported by the Air Force Office of Scientific Research Young Investigator Program under grant FA9550-19-1-0063. Work at Brookhaven National Laboratory was supported by the DOE Office of Science under Contract No. DE-SC0012704. This work was supported by the Laboratory Directed Research and Development project of Brookhaven National Laboratory No. 21-037. This work was supported by the U.S. Department of Energy (DOE) Office of Science, Early Career Research Program.  This research used beamline 2-ID of the National Synchrotron Light Source II, a U.S. Department of Energy (DOE) Office of Science User Facility operated for the DOE Office of Science by Brookhaven National Laboratory under Contract No. DE-SC0012704. This work was supported by Grant-in-Aid for Scientific Research (B) No. JP21H01804 from MEXT, Japan and by JST PRESTO grant No. JPMJPR18L2 and CREST grant No. JPMJCR16F1.

%

\newpage 

\newpage 

\newpage 

\newpage 

\newpage

\end{document}